\documentclass[useAMS,usenatbib]{mn2e}

\usepackage[dvips]{graphicx}
\usepackage{wrapfig}
\usepackage{amsmath}

\title[START: SPH with Tree-based Accelerated Radiative Transfer]
{START: Smoothed particle hydrodynamics with tree-based
accelerated radiative transfer}

\author[K. Hasegawa and M. Umemura]
{{K. Hasegawa $^{1,2}$ \thanks{E-mail: kenji.hasegawa@obspm.fr (KH)}}
and {M. Umemura $^{2}$ \thanks{E-mail: umemura@ccs.tsukuba.ac.jp(MU)}}
\\
$^{1}$LERMA, Observatoire de Paris, CNRS, 61 Av. de lfObservatoire, 75014 Paris, France\\
$^{2}$Center for Computational Sciences, University of Tsukuba, Ten-nodai, 1-1-1 Tsukuba, 
Ibaraki 305-8577, Japan}
\begin{document}

\date{Accepted 2010 May 27. Received 2010 May 23; in original form 2010 March 28}

\pagerange{\pageref{firstpage}--\pageref{lastpage}} \pubyear{2009}

\maketitle

\label{firstpage}

\begin{abstract}
We present a novel radiation hydrodynamics code, START, which is a smoothed particle hydrodynamics (SPH) scheme coupled with accelerated radiative transfer. The basic idea for the acceleration of radiative transfer is parallel to the tree algorithm that is hitherto used to speed up the gravitational force calculation in an $N$-body system. It is demonstrated that the radiative transfer calculations can be dramatically accelerated, where the computational time is scaled as $N_{\rm p} \log N_{\rm s}$ for $N_{\rm p}$ SPH particles and $N_{\rm s}$ radiation sources. Such acceleration allows us to readily include not only numerous sources but also scattering photons, even if the total number of radiation sources is comparable to that of SPH particles. Here, a test simulation is presented for a multiple source problem, where the results with START are compared to those with a radiation SPH code without tree-based acceleration. We find that the results agree well with each other if we set the tolerance parameter as $\theta_{\rm crit}\le1.0$, and then it demonstrates that START can solve radiative transfer faster without reducing the accuracy. One of important applications with START is to solve the transfer of diffuse ionizing photons, where each SPH particle is regarded as an emitter. To illustrate the competence of START, we simulate the shadowing effect by dense clumps around an ionizing source. As a result, it is found that the erosion of shadows by diffuse recombination photons can be solved. Such an effect is of great significance to reveal the cosmic reionization process.
\end{abstract}
\begin{keywords}
methods: numerical - radiative transfer - hydrodynamics - diffuse radiation
\end{keywords}

\section{Introduction}

The radiative transfer (RT) in three-dimensional space is virtually 
a six-dimensional problem for a photon distribution function 
in the phase space. 
So far, various RT schemes have been proposed, some of which
are coupled with hydrodynamics \citep{Iliev06,Iliev09}.
In a grid-based scheme, the radiative transfer can be
reduced to a five-dimensional problem without the significant
reduction of accuracy [e.g. the ART scheme that is
proposed in \citet{Nakamoto01} and \citet{Iliev06}].
In a smoothed particle hydrodynamics (SPH) scheme,
if diffuse scattering photons are neglected,
the computational cost is proportional to $N_{\rm p}N_{\rm s}$, 
where $N_{\rm p}$ and $N_{\rm s}$ are the number of SPH particles
and that of radiation sources, respectively. 
This type of radiation transfer solver can be coupled with the hydrodynamics 
(e.g. Susa 2006; Illiev et al. 2006, 2009).

But, in various astrophysical problems, diffuse scattering photons
play a significant role, for example, in photoionization of a highly
clumpy medium.
The treatment of diffuse radiation is a hard barrier owing to
its high computational cost.
Susa (2006) proposed Radiation-SPH ({RSPH}) scheme, in which the radiation
transfer based on ray-tracing is coupled with SPH. 
Since SPH particles are directly used to integrate optical depths, 
high density regions can be automatically resolved with high accuracy.
The code can actually treat multiple radiation sources, but the computational 
cost increases in proportion to the number of radiation sources. 
Hence, it is difficult to include numerous radiation sources. 
For the same reason, recombination photons are hard to involve, 
since each SPH particle should be treated as a radiation source.  
Thus, in the applications of RSPH, the on-the-spot approximation is 
assumed \citep{Spitzer78}, where recombination photons are supposed to 
be absorbed on the spot, and therefore transfer of diffuse radiation
is not solved. The RSPH has been applied to 
explore the radiative feedback on first generation object formation  
\citep{SU04, SU06, Susa07, Susa08, HUS09, SUH09},
where ionization of hydrogen and photo-dissociation of 
hydrogen molecules $\rm H_2$ by ultraviolet (UV) radiation
are key physics.
But, diffuse radiation potentially changes the ionization fraction near ionization-front (I-front),
and therefore alter $\rm H_2$ abundance.
Also, diffuse radiation can erode neutral shadows behind dense clumps.
Such erosion is significant to elucidate the cosmic reionization process.
To treat the diffuse radiation properly, the acceleration of radiation transfer
solver is indispensable. 

In this paper, we present a novel radiation transfer solver for an SPH scheme, 
START, with the tree-based acceleration of the radiative transfer. 
The paper is organized as follows. 
In Section 2, we describe the algorithm in detail. 
In Section 3, we test the new scheme by comparing with a previous scheme,
for the propagation of ionization fronts around multiple sources. 
In Section 4, we demonstrate the impacts of diffuse recombination photons 
by solving the transfer of diffuse radiation. 
Section 5 is devoted to the conclusions. Some discussion on the
related topics is also given there. 

\section{Code Description}
\subsection{Smoothed Particle Hydrodynamics}
The SPH part in our scheme basically follows \cite{Monaghan92}. 
The density is described as 
\begin{equation}
	\rho_i = \sum_j m_j W({\bf r}_{ij},h_i), 
\end{equation}
where $m_{j}$, ${\bf r}_{ij}$, $h_i$ and $W$ are the mass of $j$-th particle, the distance 
between $i$-th particle and $j$-th particle, the smoothing length of $i$-th particle, and the kernel function, respectively.  
As for the kernel function, we use the standard spline form. 
The equation of motion for each SPH particle $i$ is given by 
\begin{equation}
	\frac{d {\bf v}_i}{dt}= {\bf g}_i - \sum_j m_j \left(
	\frac{P_i}{\rho_i^2} + \frac{P_j}{\rho_j^2} + \Pi_{ij} \right)
	\nabla \bar W_{ij}, 
\end{equation}
where ${\bf g}_i$ is the gravitational acceleration and $P_{i(j)}$ is the pressure of $i(j)$-th particle. 
$\Pi_{ij}$ is the artificial viscosity for which we use the Monaghan type viscosity \citep{MG83, Monaghan92}. 
$\bar W_{ij}$ is the symmetrized kernel given by 
\begin{equation}
	\bar W_{ij} = \frac{1}{2}[W({\bf r}_{ij},h_i) + W({\bf r}_{ij},h_j)]
\end{equation} 
\citep{HK89}.
In our scheme, the gravitational force on each SPH particle is calculated by using Barnes-Hut Tree  algorithm \citep{BH86}. 

As for the equation of energy, symmetric forms have been frequently used \citep{MG83,HK89}. 
However, it is known that symmetric forms sometimes give rise to negative temperatures around a shock front \citep{Benz90}. 
Hence, we employ an asymmetric form for the equation of energy to avoid an unphysical behavior around a shock. 
Such an asymmetric form has been employed so far by several authours \citep{SM93,Umemura93,Thacker00},
where the energy equation is given by 
\begin{equation}
	\frac{du_i}{dt} = -\frac{\Lambda_i - \Gamma_i}{\rho_i} 
	+\sum_j m_j \left(\frac{P_i}{\rho_i^2} + \frac{1}{2}\Pi_{ij} \right)
	{\bf v}_{ij}\cdot \nabla_i \bar W_{ij}, 
\end{equation}
with the cooling rate $\Lambda_i$ and the heating rate $\Gamma_i$. 
This form conserves energy well \citep{Benz90,Thacker00}. In our calculations, 
the error of energy conservation is always less than 1 percent. 
The energy equation is consistently solved with the radiative transfer of UV photons 
and the non-equlibrium chemistry.

\subsection{Radiative Transfer}
\subsubsection{{RSPH} ray-tracing}

\begin{figure}
	\centering
	{\includegraphics[width=7cm]{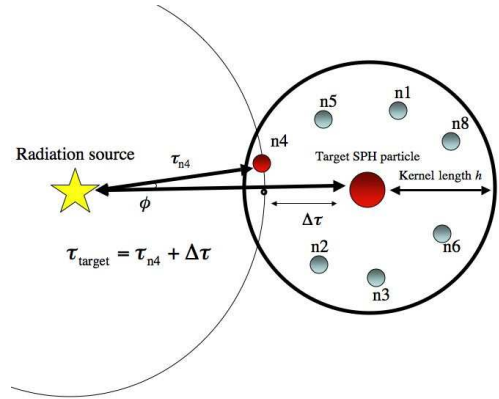}}
	\caption{The method for the ray-tracing adopted in {RSPH} scheme is schematically shown.
	The large red filled circle indicates the target particle to 
	which the optical depth from the radiation source is evaluated. 
	Smaller filled circles around the target particle indicate particles in the 
	neighbor list of the target particle.}
	\label{rsph}
\end{figure}

\begin{figure*}
	\centering
	{\includegraphics[width=17cm]{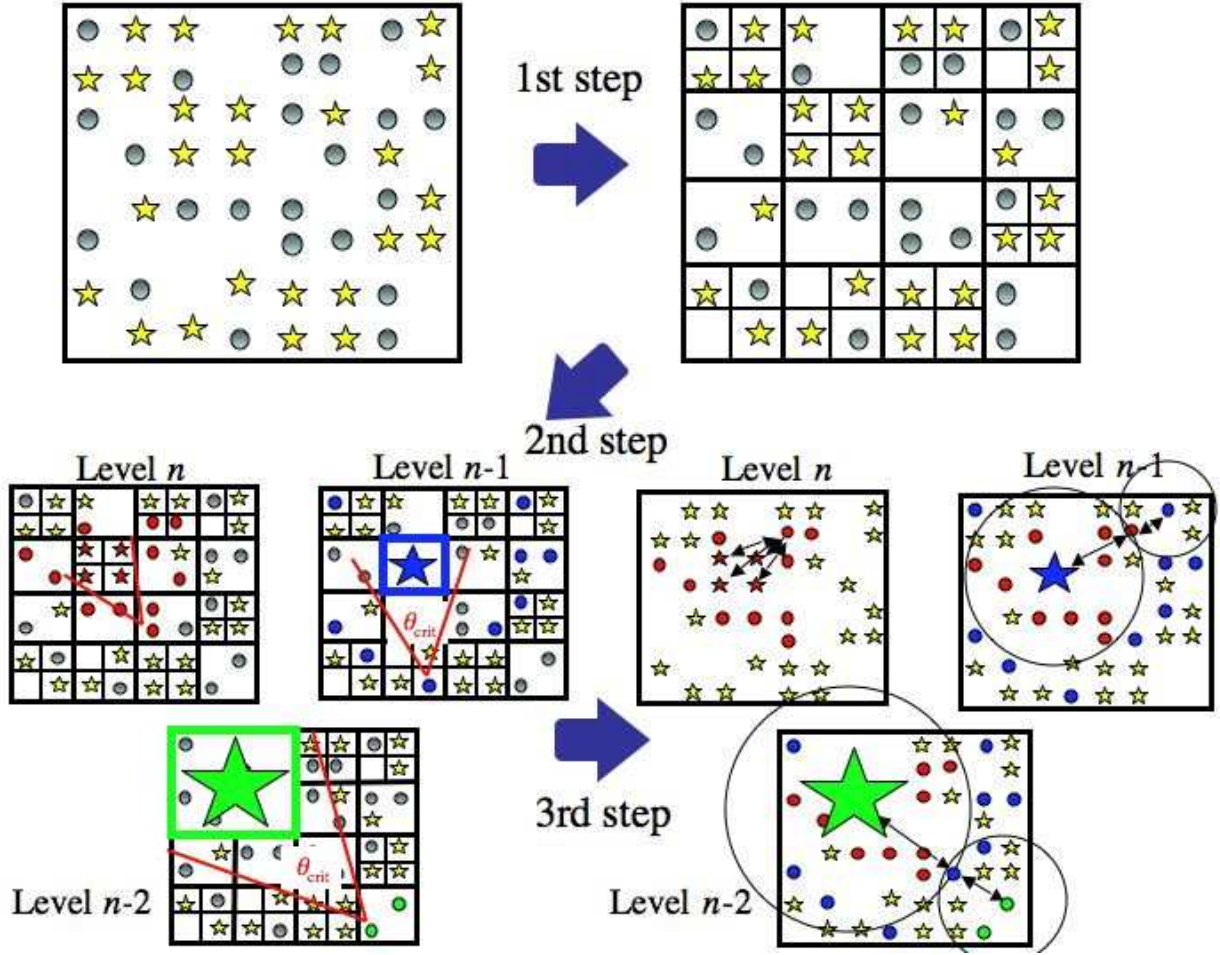}}
	\caption{The new ray-tracing method based on oct-tree structure 
	is schematically shown. 
	Each star and circle respectively indicates a radiation source and an SPH particle.
	In the first step, the tree structure for the distributions of radiation sources is constructed. 
	In the 2nd step, a list of SPH particles satisfying the same level of conditions 
	(\ref{con1}) and (\ref{con2}) is produced for each virtual radiation source. In the lower 
	right panel, particles in a list of a virtual radiation source are indicated by circles 
	filled with the same color as the owner source of the list. 
	In the 3rd step, optical depths from virtual sources to particles in the lists are evaluated in 
	the same manner as Fig. \ref{rsph}}.
	\label{schem_rsphat}
\end{figure*}

The ray-tracing algorithm in our scheme is similar to that adopted
by \cite{Susa06}, except for the tree-based acceleration that we 
newly implement.
Here, we briefly explain the RSPH scheme developed by \cite{Susa06}.

The steady radiative transfer equation is given by 
\begin{equation}
	\frac{d I_{\nu}}{d \tau_{\nu}} = -I_{\nu} + S_{\nu},
\end{equation}
where $I_{\nu}$, $\tau_{\nu}$, and $S_{\nu}$ are the specific intensity, the optical depth, and the source function, respectively.  
The equation has the formal solution given by
\begin{equation}
	I_{\nu}(\tau_{\nu}) = I_{\nu,0} {\rm e}^ {-\tau_{\nu}} + \int _0^{\tau_{\nu}}S_{\nu}(\tau_{\nu}^{\prime})
	{\rm e}^{-\tau_{\nu}+ \tau_{\nu}^{\prime}} d \tau_{\nu}^{\prime}, 
	\label{solution1}
\end{equation}
where $I_{\nu,0}$ is the specific intensity at $\tau_{\nu}=0$, and  $\tau_{\nu}^{\prime}$ 
is the optical depth at a position along the ray. 
In order to reduce the computational cost, so called on-the-spot approximation is employed \citep{Spitzer78},
where recombination photons are assumed to be absorbed on the spot.
Therefore, in this approximation, the transfer of diffuse radiation is not solved.
According to the on-the-spot approximation, equation (\ref{solution1}) is simply reduced to 
\begin{equation}
	I_{\nu}(\tau_{\nu}) = I_{\nu,0} {\rm e}^ {-\tau_{\nu}}. 
	\label{solution2}
\end{equation}
Hence, the specific intensity at a position with $\tau_{\nu}$ can be evaluated without
any iterative method for diffuse radiation.

In Fig.{\ref{rsph}}, the method for integrating an optical depth in RSPH scheme is schematically shown. 
The optical depth from the radiation source to the target SPH particle can be
approximately evaluated by 
\begin{equation}
	\tau_{\rm target} = \tau_{\rm up} + \Delta \tau,
	\label{tau}
\end{equation}
where $\tau_{\rm up}$ is the optical depth from the radiation source to 
the upstream SPH particle that is a member of neighbors of the target particle
and the closest to the light ray (in other words, having the smallest angle $\phi$).
$\Delta \tau$ is the optical depth between the target particle and 
the intersection of the light ray with a sphere which centers the the radiation source
and also includes the upstream particle (n4 in Fig.{\ref{rsph}}).
If $\tau_{\rm up}$ is evaluated in advance, we should calculate only
$\Delta \tau$. Here, $\Delta \tau$ is evaluated as 
\begin{equation}
	\Delta \tau = \sigma \Delta l \left(\frac{n_{\rm target} + n_{\rm up}}{2}\right), 
\end{equation}
where $\sigma$, $\Delta l$, $n_{\rm target}$, and $n_{\rm up}$ are the cross section, 
the distance between the intersection and the target particle, the number density 
at the position of the target particle, and the number density at the position of the upstream particle, 
respectively. 

\begin{figure*}
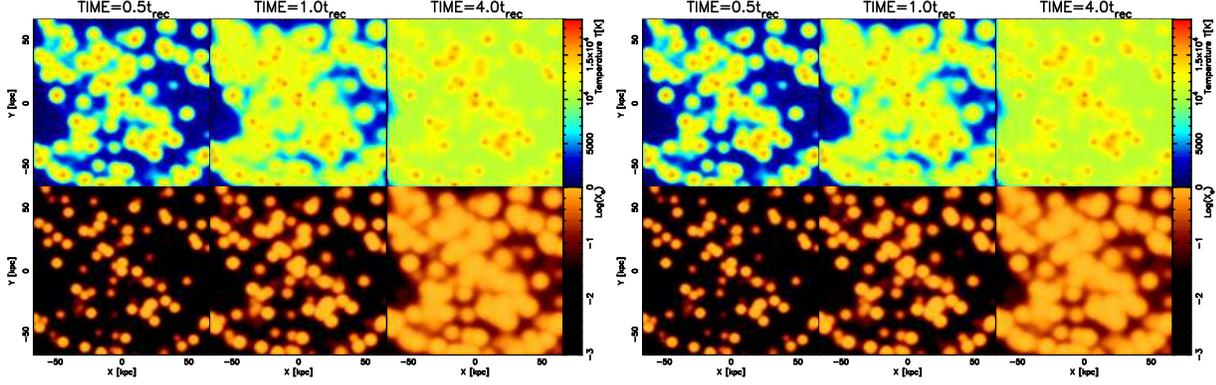

	\centering
	{\includegraphics[width=8cm]{fig3a.ps}}
	{\includegraphics[width=8cm]{fig3b.ps}}	
	\caption{{\it Test 1} -- Propagation of ionization fronts around multiple sources 
	in an optically-thick medium, which is assumed to be static and uniform.
	The hydrogen number density is $n_{\rm H}=10^{-3} {\rm cm^{-3}}$ 
	and an initial gas temperature is 100 K. 
	The simulation box size is $132{\rm kpc}$ in linear scale.	
	Upper panels show the temperature, where left three panels are
	the results by {RSPH} and right three panels are the results by {START}
	with $\theta_{\rm crit}=1$.
	Lower panels show ionization fraction in a slice through the mid-plane of the simulation box.
	Here, the results at three different times $t=0.5t_{\rm rec}$,  
	$t=1.0t_{\rm rec}$, and $t=4.0t_{\rm rec}$ are shown,  where $t_{\rm rec}$
	is the recombination time.}
	\label{test1s}
\end{figure*}

In such a method, we have to know the optical depth of the upstream particle in advance. 
If we calculate $\tau_{\rm up}$ by the integration along the light ray, 
the computational cost is roughly proportional to $N_{\rm p}^{1/3}$,
where $N_{\rm p}$ is the total number of SPH particles.
Hence, the total computational cost is
\begin{equation}
	T_{\rm calc} \propto N_{\rm p}^{1/3}N_{\rm p}N_{\rm s},	
\end{equation}
where $N_{\rm s}$ is the number of radiation sources.
But, if we utilize the optical depths for SPH particles in order of distance 
from the radiation source, only one SPH particles is used for 
the new optical depth evaluation. 
Then, the computational cost can be alleviated to
\begin{equation}
	T_{\rm calc,RSPH} \propto N_{\rm p}N_{\rm s}. 
	\label{trsph}	
\end{equation}
Nonetheless, this dependence leads to enormous computational cost,
if we consider numerous radiation sources.
Especially, to treat the diffuse scattering photons,
$N_{\rm s}$ should be equal to $N_{\rm p}$, since all SPH particles are
emission sources. Then, the computational time  
is proportional to $N_{\rm p}^2$. 
Hence, to treat diffuse radiation, further acceleration is indispensable.

\subsubsection{START: SPH with Tree-based Acceleration of Radiative Transfer}
Here, we propose a new ray-tracing method designed to 
solve the radiation transfer for numerous radiation sources.  
The procedure of the new scheme is schematically shown in Fig.\ref{schem_rsphat}. 
For simplicity, the procedure is shown in two-dimensional space, although
the scheme is implemented in three-dimensional space. 
The new scheme is based on oct-tree structure, which is widely 
used for the gravitational force calculations. 

\begin{figure}
	\centering
	{\includegraphics[width=8cm]{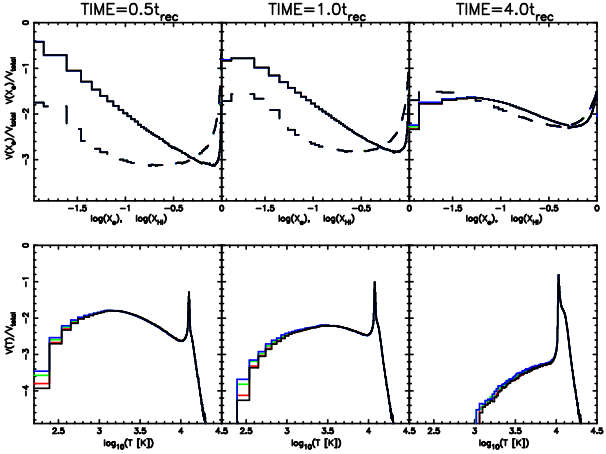}}
	\caption{Dependence on the tolerance parameter $\theta_{\rm crit}$
for Test 1. In upper panels, the volume fractions of ionized and neutral gas are 
respectively shown by solid and dashed lines, at $t=0.5t_{\rm rec}$,  $t=1.0t_{\rm rec}$, 
and $t=4.0t_{\rm rec}$. 
Black lines are the results by RSPH, while red, green, and blue lines are
the results by START with $\theta_{\rm crit}=0.6$, $\theta_{\rm crit}=1.0$, 
and $\theta_{\rm crit}=1.4$. 
Lower panels show the volume fraction of temperature.}
	\label{test1h}
\end{figure}

First, we construct the oct-tree structure for the distributions of radiation sources, 
where cubic cells are hierarchically subdivided into 8 subcells.
This procedure continues until each cell contains only one radiation source.
As the next step, we reduce the effective number of the radiation sources 
for a target particle $i$, using the oct-tree structure. 
If a cell of level $n$ is sufficiently distant from the target particle $i$, 
the radiation sources in the cell are regard as one virtual bright source 
located at the centre of luminosity in the cell. 
Practically, if the following condition is satisfied, 
the cell is regarded to contain one virtual bright source:
\begin{equation}
	\frac{l^n_{\rm cell}}{d^n} \le \theta_{\rm crit},
	\label{con1}
\end{equation}
\begin{equation}
	\frac{l^{n-1}_{\rm cell}}{d^{n-1}} > \theta_{\rm crit},
	\label{con2}
\end {equation}
where $l^n_{\rm cell}$ is the size of the $n$-th level cell and $d^n$ is the distance 
from the target particle to the closest edge of the cell, and $n-1$ indicates the parent cell of  the $n$-th level cell.
$\theta_{\rm crit}$ is the tolerance parameter, which regulates the accuracy of resulting radiation fields. 
Of course, such a judgement is done not only for $n$ and $n-1$ levels but also lower level cells, such as $n-2$, $n-3$, ..., as well. 
Therefore the effective number of radiation sources for a target particle can be dramatically reduced. 
In the present scheme, the luminosity of a virtual bright source in the $n$-th level cell is 
simply given by
\begin{equation}
	L = \sum_{j} L_{j}, 
\end{equation}
and the position is determined by
\begin{equation}
	{\bf r} = \frac{\sum_{j} L_{j}{\bf r}_{j}}{\sum_{j} L_{j}},
\end{equation}
where $L_{j}$ and ${\bf r}_{j}$ are the original luminosity and position of radiation source $j$
in the cell, respectively. 
In some situations, the evaluation with the SPH kernel as
$$
L = \frac{\sum_j L_j W({\bf r-r}_j)}{\sum_j W({\bf r-r}_j)},
$$
or 
$$
L =\sum_j \frac{m_j}{\rho_j}L_j W({\bf r-r}_j)
$$
are other possibilities. Especially, the latter may be a reasonable choice 
for the diffuse radiation for which all SPH particles are emitters. 

\begin{figure*}
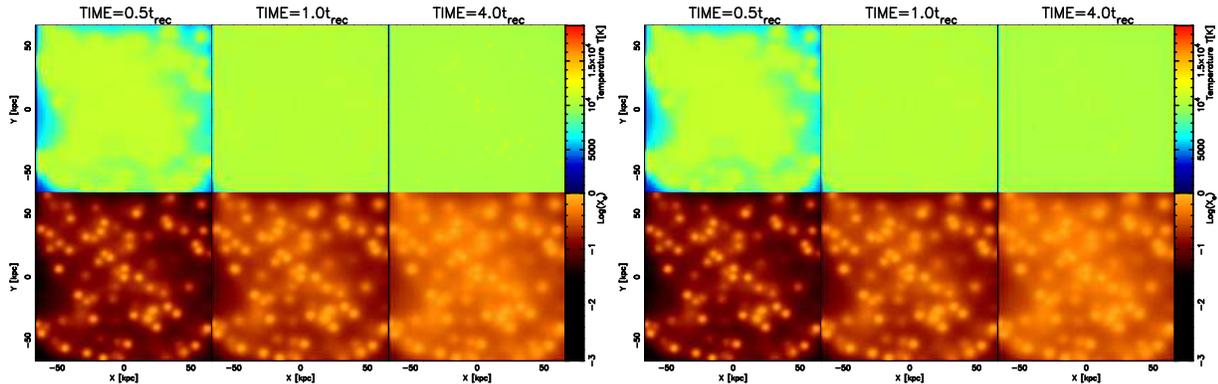

	\centering
	{\includegraphics[width=8cm]{fig5a.ps}}
	{\includegraphics[width=8cm]{fig5b.ps}}	
	\caption{{\it Test 2} -- Propagation of ionization fronts around multiple sources 
in an optically-thin medium. The same quantities as Fig. \ref{test1s} are shown.}
	\label{test2s}
\end{figure*}

Each virtual radiation source has a list of SPH particles which correspond to 
the same level of tree satisfying conditions (\ref{con1}) and (\ref{con2}). 
The list can be easily produced by using the tree structure for the distributions 
of SPH particles that has been already constructed to compute gravitational forces. 
The optical depth towards an SPH particle in the list is evaluated in order of distance 
from the radiation source by using equation (\ref{tau}). 
But, if an SPH particle is located on the boundary of the domain that the particles in the list compose,
the particle does not know $\tau_{\rm up}$. Therefore, we have to integrate the optical depth 
towards the particle along the light ray from the virtual source.

By this tree-based algorithm, 
the number of radiation sources can be dramatically reduced to 
the order of $\log N_{\rm s}$. 
Hence, with the new method, the computational cost is given by 
\begin{equation}
	T_{\rm calc, START} \propto N_{\rm p}\log N_{\rm s}. 
	\label{trsph2}
\end{equation}
Such weak dependence on $N_{\rm s}$ allows us to solve radiative transfer 
for diffuse photons emitted by all SPH particles as well as multiple sources. 
It is noted that the new ray-tracing method accords with 
{RSPH} ray-tracing in the case of $\theta_{\rm crit}=0$ or $N_{\rm s} = 1$. 

The accuracy of the new scheme depends on the value of $\theta_{\rm crit}$. 
We can easily guess that the new scheme is quite accurate in the optically thick limit, 
since the contributions of distant sources become negligible. 
In the optically thin limit, we can take $\theta_{\rm crit}$ similar 
to that used in the gravity calculation,
since the radiation flux is proportional to the gravity in this limit.
For general cases, 
we discuss what value of $\theta_{\rm crit}$ should be taken
in section 3. 

\section{Comparison between RSPH and START}
\subsection{Test calculations for the transfer of radiation from multiple sources}

\begin{figure}
	\centering
	{\includegraphics[width=8cm]{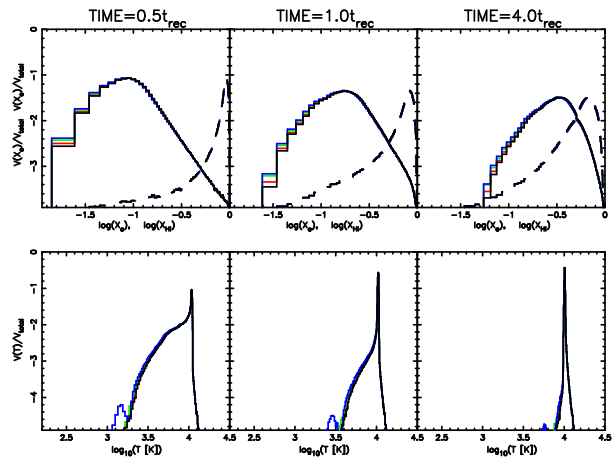}}
	\caption{Same as Fig. \ref{test1h}, except for Test 2. }
	\label{test2h}
\end{figure}

As mentioned above, the accuracy and the computational time 
depend on the tolerance parameter $\theta_{\rm crit}$ in START. 
Hence, we should chose the parameter so as to achieve reasonable accuracy
without vast computational cost. 
The best way to examine the accuracy is to compare simulation results with analytic solutions. 
Unfortunately, when multiple radiation sources exist, it is hard to obtain analytic solution.  
Therefore, we compare our results with other results simulated with a reliable scheme. 
Here, we use the results by RSPH ray-tracing method, 
since the RSPH scheme is already tested for a standard problem
and the ionization structure obtained by {RSPH} is well concordant 
with analytic solutions as shown in \cite{Iliev06} and \cite{Susa06}. 

First, we perform tests for numerous sources, 
under the on-the-spot approximation. 
We calculate the expansion of ionized regions around multiple UV sources.
In the radiation hydrodynamic process of primordial gas, not only
the ionization of hydrogen but also the photo-dissociation of 
hydrogen molecules $\rm H_2$ is important. Hence, we solve also $\rm H_2$ chemistry
(see Appendix A) and $\rm H_2$ photo-dissociation with a shielding function (see Appendix B).
We use $128^3$ SPH particles and $1,024$ radiation sources in all test calculations in this section. 

\begin{figure*}
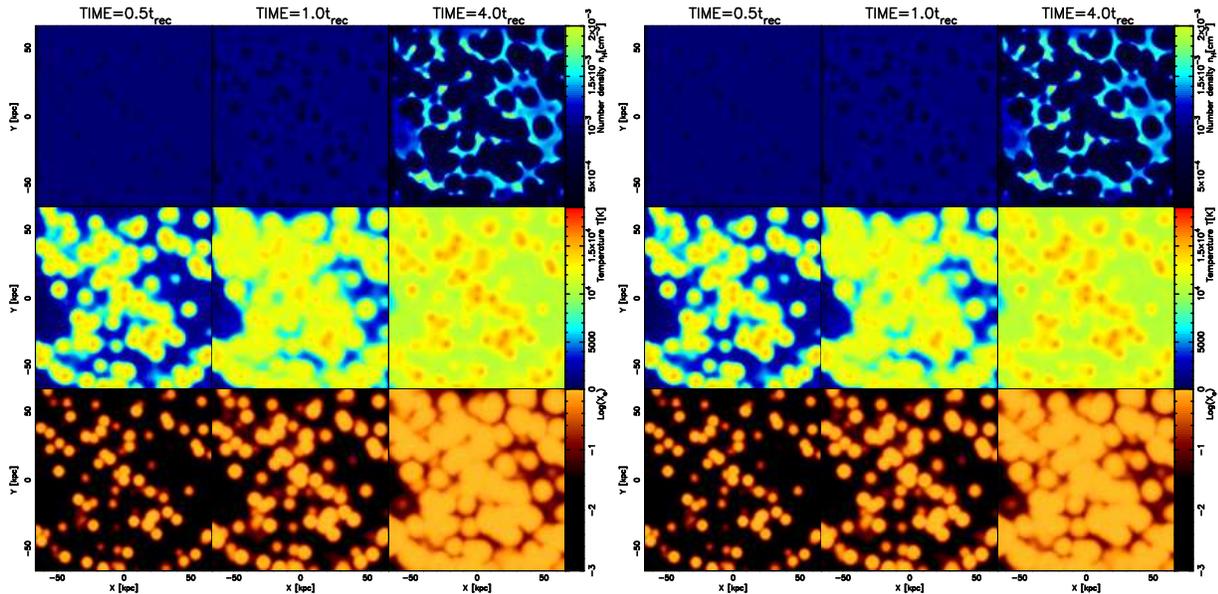

	\centering
	{\includegraphics[width=8cm]{fig7a.ps}}
	{\includegraphics[width=8cm]{fig7b.ps}}	
	\caption{{\it Test 3} -- Expansion of HII regions with hydrodynamics around
	multiple sources. The initial conditions are the same as those in Test 1.
	Upper panels depict the hydrogen number density, where left three panels are 
	the results by {RSPH} and right three panels are the results by {START}
	with $\theta_{\rm crit}=1$.
	Middle panels show the temperature.
	Lower panels show ionization fractions in a slice through the mid-plane of the simulation box.
	Here, the results at three different times $t=0.5t_{\rm rec}$,  
	$t=1.0t_{\rm rec}$, and $t=4.0t_{\rm rec}$ are shown,  where $t_{\rm rec}$
	is the recombination time.}
	\label{test3s}
\end{figure*}

\subsubsection{{\it Test 1} -- Propagation of ionization fronts in an optically-thick medium}
As the first test, we consider multiple ionizing sources 
in an neutral, uniform medum with hydrogen number density of $n_{\rm H}=10^{-3} {\rm cm^{-3}}$ 
and an initial gas temperature of 100 K. The simulation box size is $132{\rm kpc}$ in linear scale.
Here, a static medium is assumed, and therefore no hydrodynamics is solved.
The optical depth at the Lyman limit frequency initially is $\sim20$ for mean particle separation. 
The radiation sources are randomly distributed, and simultaneously radiate UV radiation when each simulation starts. 
Each UV source has the blackbody spectrum with effective temperature of $T_{\rm eff}=10^5{\rm K}$ 
and emits ionizing photons per second of $\dot N_{\gamma}=5 \times 10^{48} {\rm s^{-1}}$. 
The $\rm Str\ddot{o}mgren$ radius is analytically given by 
\begin{equation}
	R_{\rm s} = \left(\frac{3 \dot N_{\gamma}}{4\pi n_{\rm H}^2\alpha_{\rm B}}\right)^{\frac{1}{3}},
	\label{strom}
\end{equation}
where $\alpha_{\rm B}$ is the recombination coefficient to all excited levels of hydrogen. 
In the present test, the $\rm Str\ddot omgren$ radius for each radiation source 
is $5.4 {\rm kpc}$, if the HII regions around sources are not overlaped. 
Simulations are performed until the time $t=4.0t_{\rm rec}$, 
where $t_{\rm rec}$ is the recombination time defined by 
$t_{\rm rec} \equiv 1/(n_{\rm H} \alpha_{\rm B})$. 

In Fig.\ref{test1s}, the distributions of temperature and ionization fractions 
are compared between {RSPH} and {START} with $\theta_{\rm crit}=1$.
In both simulations, individual HII regions gradually expand 
and finally overlap with each other. 
As shown in the figure, the results are almost identical between {RSPH} and {START} 
at any evolutionary stage. 
In order to see the dependence on $\theta_{\rm crit}$ quantitatively, in Fig. \ref{test1h}
we plot the volume fractions of ionized and neutral components (upper panels), 
and the temperature (lower panels) for $\theta_{\rm crit}=0.6$, $\theta_{\rm crit}=1.0$, or $\theta_{\rm crit}=1.4$.
Here, the results by {RSPH} are also shown.
As for ionized and neutral gas fractions, there is little disagreement among different $\theta_{\rm crit}$.
In the volume fraction of temperature, there are a slight discrepancy only in low temperature regions. 
The discrepancy increases with increasing $\theta_{\rm crit}$. But, in ionized regions ($T>10^4$K),
the results by START agree well with {RSPH} results even for $\theta_{\rm crit}=1.4$.

\subsubsection{{\it Test 2} -- Propagation of ionization fronts in an optically-thin medium}

In the algorithm of START, a test for relatively optically-thin medium is important,
since the radiation from distant sources contribute significantly to ionization structure. 
Therefore, we carry out a test calculation, where the initial gas density is 
$n_{\rm H}=10^{-5} {\rm cm^{-3}}$, which is 100 times lower than that in Test 1. 
The mean optical depth for SPH particle separation is $\sim 0.2$. 
Each radiation source is assumed to have $5 \times \dot{N}_{\gamma}=10^{44} {\rm s^{-1}}$ 
so that the corresponding $\rm Str\ddot{o}mgren$ radius is the same as that in Test 1. 
The positions of sources are the same as those in Test 1. 

In Fig.\ref{test2s}, the distributions of temperature and ionization fractions 
are compared between {RSPH} and {START} with $\theta_{\rm crit}=1$.
Here, the early phase ($0.5t_{\rm rec}$), the expanding phase ($1.0t_{\rm rec}$), and 
the final equilibrium phase ($4.0 t_{\rm rec}$) are shown.  
Similar to Test 1, each HII region gradually expands and finally overlap each other. 
However, HII regions show more bleary structure compared to Fig.\ref{test1s}. 
Such a difference originates from the fact that the initial mean optical depth 
is 100 times smaller than that in the Test 1, and each shape of ionization front becomes vaguer. 

In Fig. \ref{test2h}, we plot the volume fractions of ionized and neutral components (upper panels), 
and the temperature (lower panels) for $\theta_{\rm crit}=0.6$, $\theta_{\rm crit}=1.0$, or $\theta_{\rm crit}=1.4$.
The results by START with $\theta_{\rm crit}=0.6$ and $\theta_{\rm crit}=1.0$ are well concordant 
with RSPH results. 
In the case of $\theta_{\rm crit}=1.4$, on the other hand, a small peak appears at several $10^3$K 
in the volume fraction of the temperature. 
Hence, $\theta_{\rm crit} \la 1$ seems to give more reliable results in this case.

\subsubsection{{\it Test 3} -- Expansion of HII regions with hydrodynamics}

\begin{figure}
	\centering
	{\includegraphics[width=8cm]{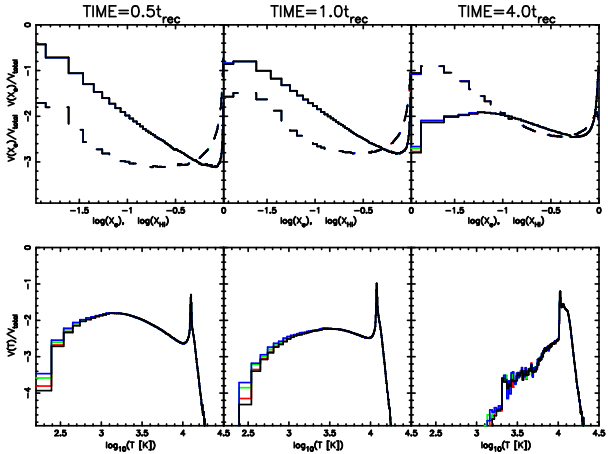}}
	\caption{Dependence on the tolerance parameter $\theta_{\rm crit}$
for Test 3. In upper panels, the volume fractions of ionized and neutral gas are 
respectively shown by solid and dashed lines, at $t=0.5t_{\rm rec}$,  $t=1.0t_{\rm rec}$, 
and $t=4.0t_{\rm rec}$. 
Black lines are the results by RSPH, while red, green, and blue lines are
the results by START with $\theta_{\rm crit}=0.6$, $\theta_{\rm crit}=1.0$, 
and $\theta_{\rm crit}=1.4$. 
Lower panels show the volume fraction of temperature.}
	\label{test3h}
\end{figure}

Here, we perform a test calculation coupled with hydrodynamics.
The radiation transfer of UV photons is consistently solved with hydrodynamics. 
The initial gas density and temperature are set to be the same as those in Test 1. 
The luminosities and positions of radiation sources are also the same as those in Test 1. 
Therefore, the difference between this test and Test 1 is solely 
whether hydrodynamics is coupled or not. 

\begin{figure}
	\centering
	{\includegraphics[width=8cm]{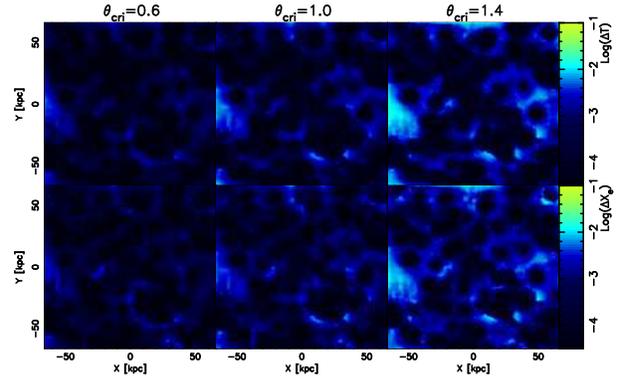}}
	\caption{Relative differences in the temperature (upper panels) and 
	the ionized gas fraction (lower panels) between START and RSPH
at the epoch $t=4.0t_{\rm rec}$. 
	From left to right column, the relative differences between 
	RSPH and START with $\theta_{\rm crit}=0.6$, $\theta_{\rm crit}=1.0$,  
	and $\theta_{\rm crit}=1.4$ are shown.}
	\label{test3e}
\end{figure}

In Fig.\ref{test3s}, the results by {START} with $\theta_{\rm crit}=1.0$ are compared
with the results by {RSPH} in terms of hydrogen number density, temperature,
and ionization fractions.
As shown in this figure, there is no distinct difference between RSPH and START, 
even if we include hydrodynamics. 
In Fig.\ref{test3h}, we show the dependence on $\theta_{\rm crit}$
in terms of volume fractions of ionized and neutral components (upper panels), 
and temperature (lower panels) at three different phases $0.5t_{\rm rec}$, $1.0t_{\rm rec}$, and  $4.0t_{\rm rec}$. 
As shown in the figure, the volume fractions for all simulations look well concordant. 
In Fig.\ref{test3e}, we show relative differences in temperature (upper panels) 
and ionized gas fraction (lower panels) between START and RSPH.
Here, three START simulations with $\theta_{\rm crit}=0.6$, $\theta_{\rm crit}=1.0$, and
$\theta_{\rm crit}=1.4$ are compared with RSPH at $t=4.0t_{\rm rec}$.
The distributions are given for the mid-plane of the computational box.   
The relative difference of a physical quantity is evaluated as
\begin{equation}
	\Delta Q = \frac{|Q_{\rm START} - Q_{\rm RSPH}|}{Q_{\rm RSPH}},
	\label{relative}
\end{equation}
where $Q_{\rm START}$ and $Q_{\rm RSPH}$ are the quantities obtained by START and by RSPH, respectively. 

As seen in Fig.\ref{test3e}, 
the relative differences increases according as the tolerance parameter becomes larger. 
Moreover, the relative differences in the neutral regions are higher than 
those in the highly ionized region, since the physical quantities in such neutral regions 
are sensitive to the incident ionizing flux. 
Fig.\ref{test3e} shows that if we choose the tolerance parameter of $\theta_{\rm crit}\la 1.0$,
the relative differences become as small as $\la 10^{-2}$ everywhere. 

\subsection{Computational time}
\begin{figure}
	\centering
	{\includegraphics[width=8cm]{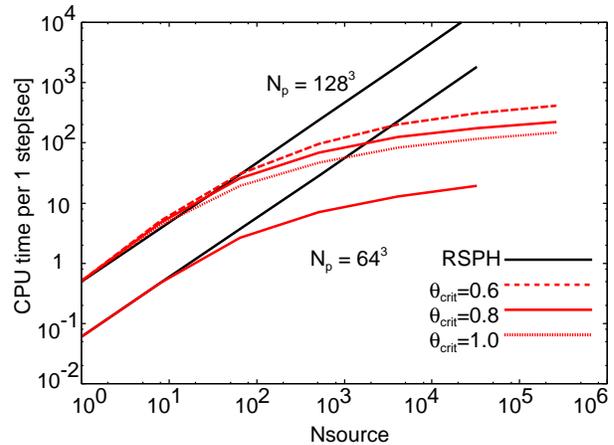}}
	\caption{Average computational time per one step 
	is shown as a function of number of radiation sources
	for SPH particles of $N_{\rm p}=128^3$ and $N_{\rm p}=64^3$. 
	The computational time of START and RSPH are denoted by red
	and black lines, respectively. 
	As for START, the calculations with 
	$\theta_{\rm crit}=0.6$, $0.8$, and $1.0$ are indicated by dashed, 
	solid, and dotted lines, respectively.}
	\label{time1}
\end{figure}
In this section, we show how radiative transfer calculations are accelerated 
by the new ray-tracing method. 
The calculations have been done with one Xeon processor (12Gflops). 
The calculation time is measured in Test 3 with varying the number of radiation sources. 
The calculations are done until $t=1.0t_{\rm rec}$ with $\sim 30-40$ time steps. 
In Fig.\ref{time1}, the mean computational time per step is shown 
as a function of the radiation source number. 
In this figure, the computational time by RSPH is denoted by black line, 
while that by START with $\theta_{\rm crit}=0.6$, 0.8, and 1.0 are denoted 
by red dashed, solid, and dotted lines, respectively. 
The SPH particle number is $N_{\rm p}=128^3$ or $N_{\rm p}=64^3$. 
In the case with $N_{\rm p}=64^3$, START is always faster than RSPH. 
This actually comes from the reduction of the effective number of radiation sources. 
In the case of $N_{\rm p}=128^3$, START is faster than RSPH as long as
$N_{\rm s}>10$. 
The calculation with $N_{\rm s}=1,024$ by START is roughly 30 times faster than that by RSPH,
and we find that the results obtained by START are well concordant with those by RSPH.
The START allows the acceleration by $N_{\rm p}\log N_{\rm s}$ for a large number of
sources, as anticipated.
However, START is a bit slower than {RSPH} for $N_{\rm s}<10$. 
In the START scheme, each radiation source (cell) has a list of SPH particles, 
to which optical depths are evaluated.
When we calculate the optical depths,  we also have to evaluate optical depths 
to some SPH particles that are not in the list. 
The time for calculating this additional part increases according as the number of SPH particles increases. 
In addition, the effective source number is hardly reduced in the case of $N_{\rm s}\sim10$. 
Such an increase of computational time is noticeable for $N_{\rm p}=128^3$ run, but not in $N_{\rm p}=64^3$ run. 
Anyway, the acceleration by START allows us to treat a large number of sources. 

\section{Radiative transfer of diffuse radiation}

In the previous sections, we have shown the acceleration by START for multiple radiation sources.
Here, we present the effectiveness of START when we include diffuse radiation, where
all SPH particles should be emitting sources. We concentrate on the photoionization problem again
and include diffuse recombination photons.

\subsection{Physical process}

To consider the transfer of recombination photons, we adopt the recombination coefficient 
to all bound levels of hydrogen $\alpha_{\rm A}(T)$, instead of that to all excited levels of hydrogen $\alpha_{\rm B}(T)$
in the on-the-spot approximation. 
We solve the transfer of recombination photons from each SPH particle 
to all other particles by using equation (\ref{solution2}) and oct-tree structure. 
The number of recombination photons emitted per unit time by an SPH particle $j$ 
is expressed by
\begin{equation}
	\dot {N}_{{\rm rec},j}= \alpha_{1}(T_{j}) n_{{\rm e,} j} n_{{\rm p},j} V_j, 
\end{equation} 
where $T_{j}$, $n_{\rm e}$, $n_{\rm p}$, and $V_{j}$ are the temperature, 
the electron number density, the proton number density, and the volume of $j$-the particle, respectively.  
$\alpha_1(T)$ is the recombination coefficient to the ground state of hydrogen, i.e.,  
$\alpha_1(T)=\alpha_{\rm A}(T) - \alpha_{\rm B}(T)$. 
The emitted energy per one recombination is $\sim h \nu_{\rm L} + kT_{j}$, 
where $h$ and $k$ are respectively the Planck constant and the Boltzmann constant. 
The temperature of photoionized gas is typically $\sim 10^4$ \citep[e.g.,][]{UI84,TW96}, 
and therefore $h\nu_{\rm L} \gg kT_{j}$.  
Thus, the number of recombination photons emitted by particle $j$ 
per unit time and solid angle is approximately given by 
\begin{equation}
	\dot {n}_{\nu,j} \approx \left\{
	\begin{array}{cc}
	\dot {N}_{{\rm rec},j} /4\pi \delta \nu
~~~~~~~~~~~~~~~~\cdots~ \nu_{\rm L} \le \nu \le \nu_{\rm L} + \delta \nu, &\\
	0~~~~~~~~~~~~~~~~~~~~~~~~\cdots~{\rm otherwise},  &
	\end{array}
	\right. 
\end{equation}
where $\delta \nu = kT_{j}/h$. 
As described in Section 2, if a cell is far enough from a target particle, 
SPH particles emitting recombination photons in the cell are regarded as one bright emitter. 
The position and the recombination photon number of the bright source labeled 
$\alpha$ are respectively determined by 
\begin{equation}
	{\bf r_{\alpha}} = \frac{\sum_j \dot {N}_{{\rm rec},j}{\bf r}_j}{\sum_j \dot {N}_{{\rm rec},j}},
\end{equation}
\begin{equation}
	{ \dot{n}_{\nu ,\alpha}} = \frac{\sum_j \dot{N}_{{\rm rec},j}}{4\pi \delta \bar {\nu}}. 
	\label{ssum}
\end{equation}
Here $\delta \bar{\nu}$ is defined by 
\begin{equation}
	\delta \bar{\nu} = \frac{k}{h}\frac{\sum_j \dot{N}_{{\rm rec},j} T_j}{\sum_j \dot{N}_{{\rm rec},j}}. 
\end{equation}
As for the recombination photon number for virtual sources, 
using another form by the SPH kernel interpolation instead of  equation (\ref{ssum})
might be a reasonable way. 
Using equation (\ref{solution2}) instead of equation (\ref{solution1}), 
the photoionization rate caused by $\alpha$-th radiation source on $i$-th particle is given by 
\begin{equation}
	k_{\alpha} = n_{{\rm HI}}({\bf r}_i) \int \int_{\nu_{\rm L}}^{\nu_{\rm L} + \delta \bar{\nu}} \dot{n}_{\nu,\alpha} 
		{\rm e}^{-\tau_{\nu,\alpha}}\sigma_{\nu} d\nu d\Omega. 
\end{equation}
Here the integration with respect to solid angles is done by considering the effect of geometrical dilution. 
The total photoionization rate at ${\bf r}_i$ is given by
\begin{equation}
	k_{\rm ion} = k_{\rm self} + k_{\rm other}, 
	\label{ktot}
\end{equation}
where $k_{\rm self}$ is the contribution from $i$-th particle itself and
$k_{{\rm other}}$ is the contributions from all emitting particles except for $i$-the particle.
$k_{{\rm other}}$ is given by   
\begin{equation}
	k_{{\rm other}} = \sum_{\alpha}k_{\alpha}.
	\label{ksum} 
\end{equation}
Applying equation (\ref{solution1}) and assuming the isotropic radiation field around $i$-th particle, 
$k_{\rm self}$ is given by 
\begin{equation}
	k_{\rm self} = 4\pi n_{\rm HI} \int_{\nu_{\rm L}}^{\nu_{\rm L} + \delta_{\nu}}
	\int_{0}^{\tau_{\nu}} \frac{S_{\nu}}{h\nu} \sigma_{\nu} 
	{\rm e}^{-\tau_{\nu} + \tau_{\nu}^{\prime}} d\tau_{\nu}^{\prime} d\nu, 
	\label{kself1}
\end{equation}
where $\tau_{{\nu}}$ is the optical depth from the edge to the centre of the particle, which is given by
\begin{equation}
	\tau_{\nu} = \sigma_{\nu} n_{\rm HI} \left(\frac{3 V}{4\pi}\right)^{1/3}.
\end{equation} 
In addition, the source function $S_{\nu}$ can be provided by
\begin{equation}
	S_{\nu} = \left\{
	\begin{array}{cc}
	\frac{h\nu \alpha_{1}(T) n_{\rm e} n_{\rm p}}
	{4\pi n_{\rm HI} \sigma_{\nu} \delta \nu }
	~~~~\cdots~\nu_{\rm L} \le \nu \le \nu_{\rm L} + \delta \nu, &\\
	0~~~~~~~~~~~~~~~~~~~~\cdots~{\rm otherwise},  &
	\end{array}
	\right. 
\end{equation}
Here the source function can be regarded as almost constant near the centre of particle $i$. 
As a result, equation (\ref {kself1}) can simply be given by 
\begin{equation}
	k_{\rm self} = \int_{\nu_{\rm L}} ^{\nu_{\rm L} + \delta_{\nu}}
	\frac{\alpha_{1}(T)}{\delta \nu} n_{\rm e} n_{\rm p}
	\left(1-{\rm e}^{-\tau_{\nu}} \right) d\nu. 
\end{equation}
Note that equation (\ref{ktot}) accords with $\alpha_{1}(T) n_{\rm e} n_{\rm p}$, 
if the medium is very opaque everywhere. 
It means that the 'net' recombination coefficient corresponds to $\alpha_{\rm B}$ in that case. 
We also evaluate the photoheating rate caused by the recombination photons in a similar way. 
The photoheating due to the recombination photons emitted by $i$-th particle itself, 
and by all other particles are respectively given by 
\begin{equation}
	\Gamma_{\rm self} = \int_{\nu_{\rm L}} ^{\nu_{\rm L} + \delta_{\nu}}
	\frac{\alpha_{1}(T)}{\delta \nu} n_{\rm e} n_{\rm H^+} 
	h(\nu - \nu_{\rm L})\left(1-{\rm e}^{-\tau_{\nu}} \right) d\nu, 
	\label{gself}
\end{equation}
and
\begin{equation}
	\begin{array}{lll}
	\Gamma_{\rm other} &=& n_{\rm HI} \sum_{\alpha} \int 
	\int_{\nu_{\rm L}}^{\nu_{\rm L} + \delta \bar{\nu}} \dot{n}_{\nu,\alpha} \\
	&&\times h(\nu - \nu_{\rm L}) {\rm e}^{-\tau_{\nu,\alpha}}\sigma_{\nu} d\nu d\Omega. 
	 \end{array}
	\label{gother}
\end{equation}
Finally, the total photoheating rate due to the recombination radiation for each SPH particle is 
given by 
\begin{equation}
	\Gamma_{\rm heat} = \Gamma_{\rm self} + \Gamma_{\rm other}.
	\label{gtot}
\end{equation}

\subsection{Test calculations for the transfer of recombination photons}

In this section, we explore the effect of recombination photons 
by performing some test calculations.  
All simulations in this section are performed with $128^8$ SPH particles. 
The transfer of recombination photons is solved by START with $\theta_{\rm crit}=0.8$. 

\subsubsection{Test 4 -- Expansion of an HII region in a static medium}

\begin{figure}
	\centering
	{\includegraphics[width=8cm]{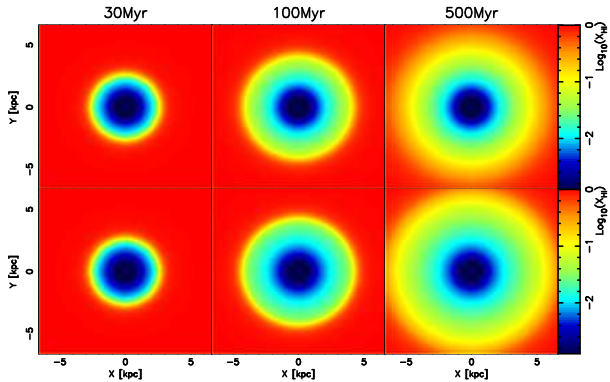}}
	\caption{{\it Test 4} -- Expansion of an HII region in a static medium.
	Upper three panels are the results by the on-the-spot approximation, while
	lower three panels are the results by solving the transfer of diffuse recombination photons.
The neutral fractions in a slice through the mid-plane of the 
	computational box are shown at $t=30 {\rm Myr}$ (left panels), $t=100 {\rm Myr}$ (middle panels), 
	and $t=500 {\rm Myr}$ (right panels). 
}
	\label{test4X}
\end{figure}
\begin{figure}
	\centering
	{\includegraphics[width=8cm]{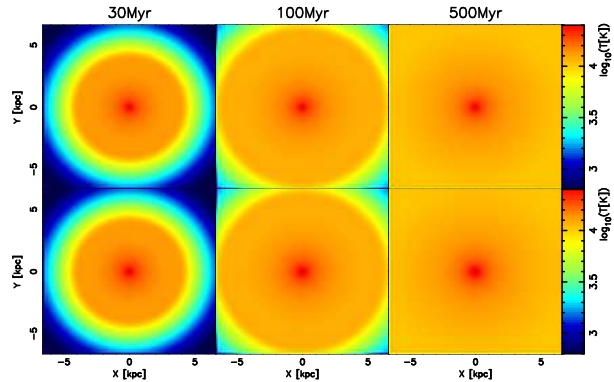}}
	\caption{The temperature distributions in Test 4. }
	\label{test4T}
\end{figure}

In this test, we simulate an expansion of an HII region in a uniform static medium.  
The initial physical parameters of this test are the same as those of Test 2 
in Cosmological Radiative Transfer Comparison Project \citep{Iliev06}, where 
the hydrogen number density
is $n_{\rm H} = 10^{-3} {\rm cm^{-3}}$, the gas temperature is $T=10^2{\rm K}$, 
and the ionization fraction is $1.2 \times 10^{-3}$. 
The computational box is 13.2 kpc in linear scale. 
One radiation source with an effective temperature $T_{\rm eff} = 10^5{\rm K}$ 
and the number of ionizing  photons per unit time 
as $\dot {N}_{\gamma}=5\times 10^{48} {\rm s^{-1}}$ 
is set up at the centre of the computational box ([$x$, $y$, $z$]=[0, 0, 0] kpc). 
The ${\rm Strom\ddot{o}mgren}$ radius estimated by equation (\ref{strom}) is 5.4 kpc. 
The simulation is performed until 500 Myr, which roughly corresponds to $4{t_{\rm rec}}$. 

In Fig.\ref{test4X}, 
we show the neutral fractions in a slice through the mid-plane of the computational box 
at a fast expanding phase ($t=30 {\rm Myr}$), a slowing-down phase ($t=100 {\rm Myr}$), 
and the final equilibrium phase ($t=500 {\rm Myr}$) . 
Fig.\ref{test4T} gives the temperature distributions. 
In these figures, we compare the results by solving the transfer of recombination photons 
with those by on-the-spot approximation.
Although the difference between these results is not large at $t=30 {\rm Myr}$ or $t=100 {\rm Myr}$,
the difference is distinct at later phase at the final equilibrium phase ($t=500 {\rm Myr}$).
The ionized region simulated with the transfer of recombination photons is more extended 
than that simulated with the on-the-spot approximation. 

\begin{figure}
	\centering
	{\includegraphics[width=8cm]{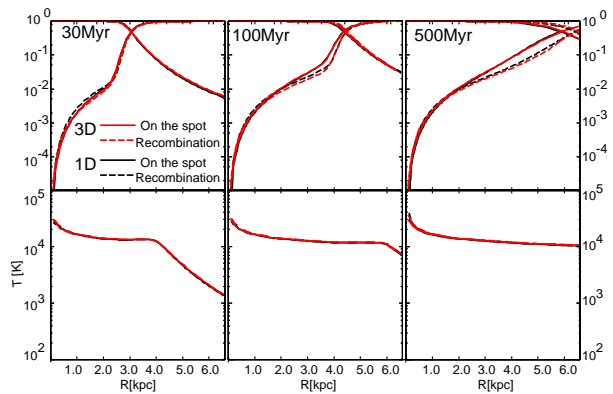}}
	\caption{For Test 4, radial profiles of the ionized fraction, the neutral fraction (upper row), and 
	the temperature (lower row) at $t=30 {\rm Myr}$, $t=100 {\rm Myr}$, and $t=500{\rm Myr}$ 
	are shown. The profiles obtained by three-dimensional radiation transfer START, and 
	one-dimensional 
	spherical symmetric radiation transfer are indicated by red and black lines, respectively. 
	Solid lines show the results by the on-the-spot approximation, while
	dashed lines show those by the full transfer of recombination photons.}
	\label{test4p}
\end{figure}

In order to confirm whether the transfer of recombination photons is correctly solved, 
we compare the results obtained by START with those by a one-dimensional spherical 
symmetric RHD code developed by \cite{Kitayama04} 
in which the transfer of recombination photons is solved. 
In 1D RHD simulations, 600 gas shells are used.
In Fig.\ref{test4p}, we show the radial profiles of the ionized fraction, the neutral fraction, 
and the temperature at each evolutionary phase. 
As shown in this figure, the results of the 3D-RHD simulations and of the 1D RHD simulation 
are well consistent with each other. 
The difference between the profiles in the full transfer and the on-the-spot approximation 
appears in moderate regions in which the neutral fractions are $x \sim 0.1$. 
In these regions, the ionized fractions calculated by solving the full transfer 
are slightly higher than those calculated by assuming the on-the-spot approximation. 
On the other hand, near the radiation source, the neutral fractions obtained 
by the full transfer are slightly lower, in contrast to the moderate ionized regions. 
This originates from the fact that $\alpha_{\rm A}$ is greater than $\alpha_{\rm B}$. 

\subsubsection{{\it Test 5} -- Expansion of an HII region with hydrodynamics}

\begin{figure}
	\centering
	{\includegraphics[width=8cm]{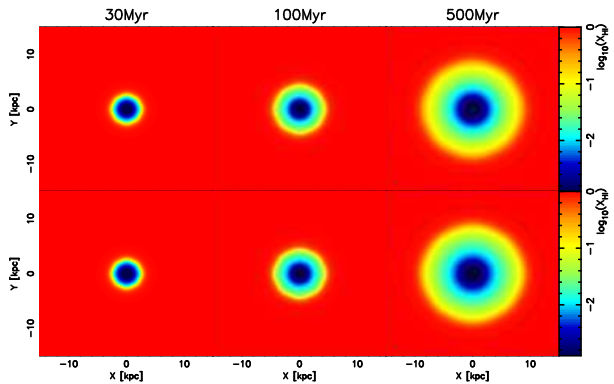}}
	\caption{{\it Test 5} -- Expansion of an HII region with hydrodynamics.
	Upper three panels are the results by the on-the-spot approximation, while
	lower three panels are the results by solving the transfer of diffuse recombination photons.
The neutral fractions in a slice through the mid-plane of the 
	computational box at $t=30 {\rm Myr}$ (left panels), $t=100 {\rm Myr}$ (middle panels), 
	and $t=500 {\rm Myr}$ (right panels) are shown.}
	\label{test5X}
\end{figure}

\begin{figure}
	\centering
	{\includegraphics[width=8cm]{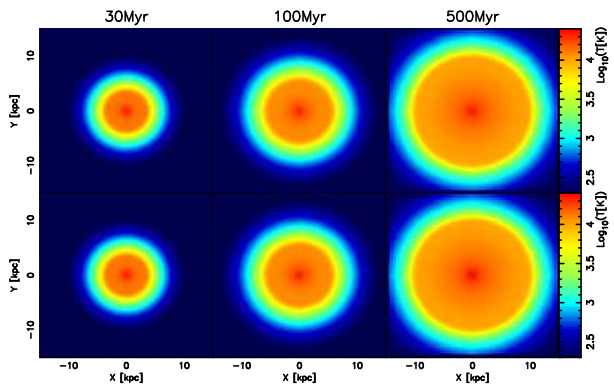}}
	\caption{The temperature distributions in Test 5. }
	\label{test5T}
\end{figure}

\begin{figure}
	\centering
	{\includegraphics[width=8cm]{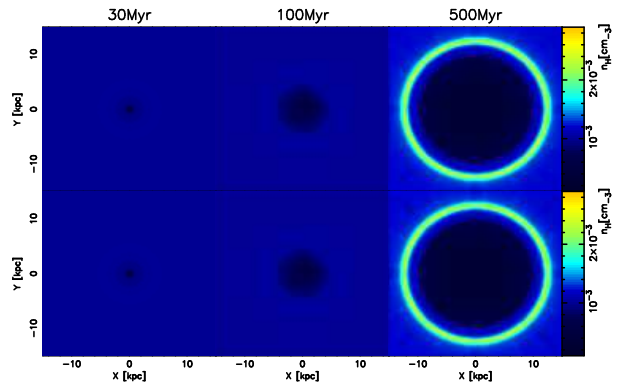}}
	\caption{The hydrogen number density distributions in Test 5. }
	\label{test5rho}
\end{figure}

This test is similar to Test 4, but here hydrodynamics is consistently solved with the radiative transfer. 
The initial physical parameter set of the test is the same as those of Test 5 
in Cosmological Radiative Transfer Comparison Project \citep{Iliev09},
where the initial gas density and temperature are $10^{-3}{\rm cm^{-3}}$ and $100 {\rm K}$. respectively. 
The size of the computational box is 30 kpc in linear scale. 
A radiation source with $T_{\rm eff} = 10^5{\rm K}$ and $\dot{N}_{\gamma}= 5\times 10^{48} {\rm s^{-1}}$ 
is set at the centre of the computational box ([$x$, $y$, $z$]=[0, 0, 0] kpc). 

The neutral fractions, the temperature, and the hydrogen number density 
in a slice through the mid-plane of the computational box are shown 
in Figs.\ref{test5X}, \ref{test5T}, and \ref{test5rho}, respectively. 
In these figures, we show the distributions of these quantities at three different evolutionary phases. 
In the first phase, R-type ionization front (I-front) rapidly propagates ($t=30{\rm Myr}$). 
In the second phase, the transition of I-front from R-type to D-type occurs ($t=100{\rm Myr}$), 
and finally in the third phase, D-type I-front preceded by a shock propagates ($t=500{\rm Myr}$). 
These figures show that if hydrodynamics is coupled, the resultant ionization structure
is in a quite good agreement with the results by the on-the-spot approximation.

\begin{figure}
	\centering
	{\includegraphics[width=8cm]{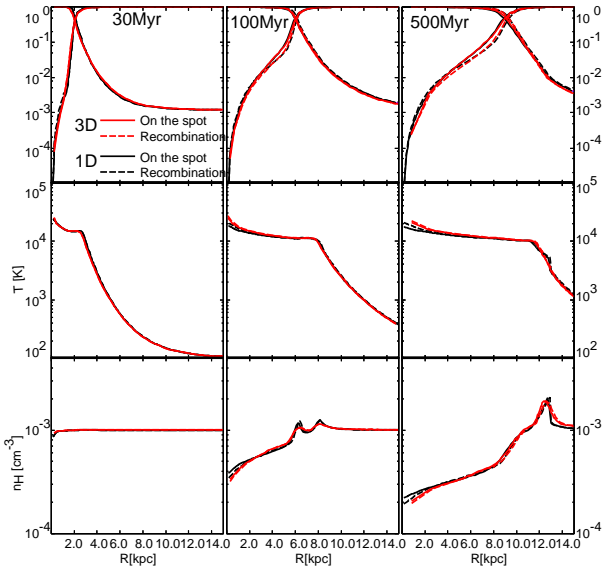}}
	\caption{For Test 5, radial profiles of the ionized fraction, the neutral fraction (top row),  
	the temperature (middle row), and the hydrogen number density (bottom row) 
	at $t=10 {\rm Myr}$, $t=200 {\rm Myr}$, and $t=500{\rm Myr}$ 
	are shown. The profiles obtained by three-dimensional radiation transfer, and one-dimensional 
	spherical symmetric radiation transfer are indicated by red and black lines, respectively. 
Solid lines show the results by the on-the-spot approximation, while
dashed lines show those by the full transfer of recombination photons.}
	\label{test5p}
\end{figure}

In Fig,\ref{test5p}, we compare the results with a 1D-RHD simulation.
Here, the averaged radial profiles of the ionized fraction, the neutral fraction, 
the temperature, and the hydrogen number density are shown
at three different times $t=10 {\rm Myr}$, $t=200 {\rm Myr}$, and $t=500{\rm Myr}$. 
At every phase, the profiles in 3D-RHD simulations are well concordant with 1D-RHD results.
Similar to the case without hydrodynamics, it is found that the recombination photons 
slightly change the profile of ionized fraction near the I-front. 
On the other hand, the profile of the temperature is not strongly affected by the recombination photons. 
As a result, the gas dynamics does not change dramatically in this case, 
even if we consider the transfer of recombination photons.

\subsubsection{{\it Test 6} -- Shadowing effects by dense clumps}

\begin{figure}
	\centering
	{\includegraphics[width=8cm]{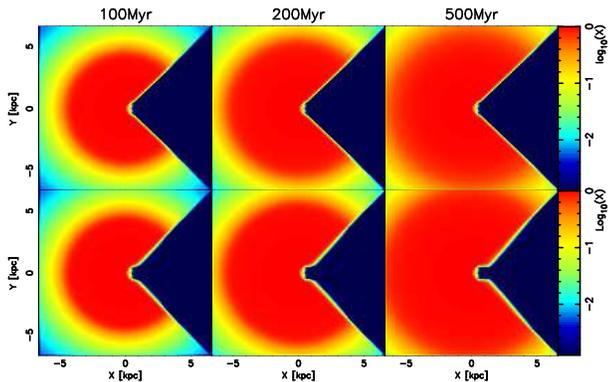}}
	\caption{{\it Test 6} -- Shadowing effects by a dense clump.
The hydrogen number density of the clump and surrounding medium 
is respectively $n_{\rm c}=2.0\times 10^{-1}{\rm cm^{-3}}$ and
$n_{\rm H} = 10^{-3} {\rm cm^{-3}}$. The clump radius is $r_{\rm c}=0.56$kpc.
	Upper three panels are the results by the on-the-spot approximation, while
	lower three panels are the results by solving the transfer of diffuse recombination photons.
The neutral fraction in a slice through the mid-plane of the 
	computational box are shown at $t=100 {\rm Myr}$ (left panels), $t=200 {\rm Myr}$ (middle panels), 
	and $t=500 {\rm Myr}$ (right panels). }
	\label{test6Xd}
\end{figure}

\begin{figure}
	\centering
	{\includegraphics[width=8cm]{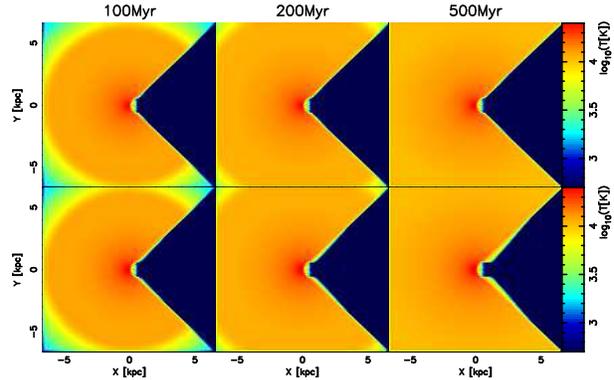}}
	\caption{The temperature distributions in Test 6. }
	\label{test6Td}
\end{figure}

Here, we explore the shadowing effect behind dense clumps. 
The physical parameters in this test is the same as those in Test 4, 
except that a dense clump with the radius of $r_{\rm c}=0.56$kpc
is placed at 0.8 kpc away in the $x$-direction from the box centre ([0.8, 0, 0] kpc).  
The hydrogen number density of the clump is $n_{\rm c}=2.0\times 10^{-1}{\rm cm^{-3}}$, 
which is 200 times higher than that in the surrounding medium
of $n_{\rm H} = 10^{-3} {\rm cm^{-3}}$. 
The ${\rm Str\ddot{o}mgren}$ radius $r_{\rm s}$ for the density of clump 
is given by 
\begin{equation}
	r_{\rm s} = \frac{F}{\alpha_{\rm B} n_{\rm H}^2}, 
\end{equation}
where $F$ is the incident photon number flux. 
Assuming $T=10^4{\rm K}$ and using $F$ at the centre of the clump, 
the $\rm Str\ddot{o}mgren$ radius is roughly $r_{\rm s} \approx 0.025 {\rm kpc}$. 
Since $r_{\rm c} \gg r_{\rm s}$, 
the clump is readily shielded from the incident radiation. 
Thus, it is expected that the ionization front is trapped in the clump 
and a shadow is created behind the clump. 

In Figs.\ref{test6Xd} and \ref{test6Td}, 
we show the resultant ionized fraction and temperature 
in a slice through the mid-plane of the box. 
In each figure, the upper panels show the result by on-the-spot approximation, 
while the lower panels show that by the transfer of recombination photons. 
The ionization front is expectedly trapped in the dense clump and a shadow is created.
Under the on-the-spot approximation, only the transfer of UV photons 
along the radial direction from the radiation source is included. 
On the other hand, when we solve the transfer of diffuse recombination photons, 
the recombination photons from the other directions photoinize the shadow. 
We can observe such an erosion of shadow in Figs. \ref{test6Xd} and \ref{test6Td}. 
But, the erosion is not so drastic in this case. It is related to the mean free path of ionizing photons.
The mean free path of ionizing photons at Lyman limit frequency $L_{\rm mfp}$ is given by 
\begin{equation}
	L_{\rm mfp} = 51.4\times \left( \frac{10^{-3}{\rm cm^{-3}}}{n_{\rm H}}\right) {\rm pc}.  
	\label{mfp}
\end{equation}
This implies that the initial mean free path in surrounding low density regions 
is approximately one tenth of the size of clump. The size of erosion is just
corresponding to the the mean free path of ionizing photons. 
That means that in lower density of surrounding medium, the erosion is more significant.

\begin{figure}
	\centering
	{\includegraphics[width=8cm]{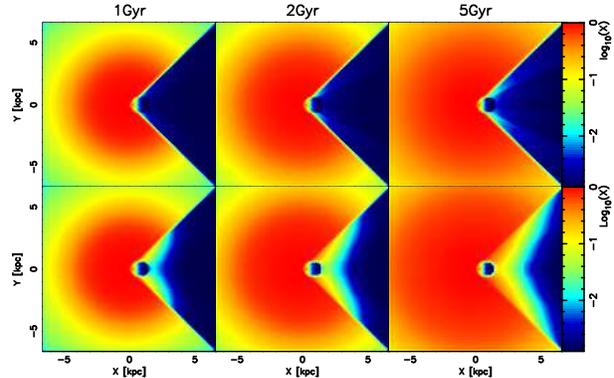}}
	\caption{Same as Fig.\ref{test6Xd}, but for the lower density of
surrounding medium as $n_{\rm H} = 10^{-4} {\rm cm^{-3}}$.}
	\label{test6Xr}
\end{figure}
\begin{figure}
	\centering
	{\includegraphics[width=8cm]{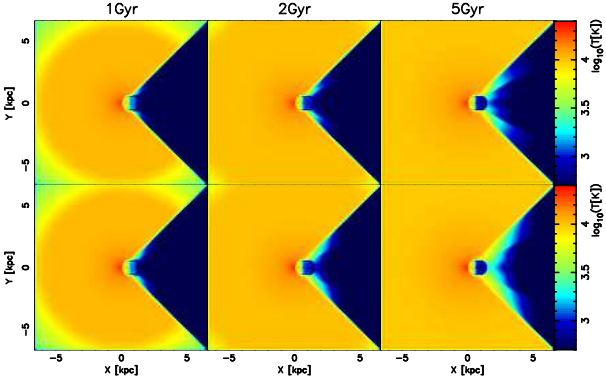}}
	\caption{The temperature distributions in the case of Fig.\ref{test6Xr}.}
	\label{test6Tr}
\end{figure}

In order to confirm the effect of the mean free path, we simulate the case in which
the number density of surrounding medium is one tenth of that in previous case. 
In this case, the mean free path is roughly comparable with the size of the clump. d
We show the ionized fraction in a slice through the mid-plane of the box in Fig.\ref{test6Xr}. 
The results by the on-the-spot approximation and those by
the full transfer are shown in the upper panels and the lower panels, respectively.   
As shown in the figure, the recombination photons can gradually ionize the gas 
behind the clump, in contrast to the higher density case. 
We show the temperature distributions in Fig.\ref{test6Tr}. 
Owing to photoheating by the recombination photons, the gas behind the clump is heated up. 
Since the number density assumed here roughly corresponds to that 
in intergalactic medium (IGM) at high redshifts, it is expected that the recombination 
photons play an important role for the reionization of the Universe.  

\begin{figure}
	\centering
	{\includegraphics[width=8cm]{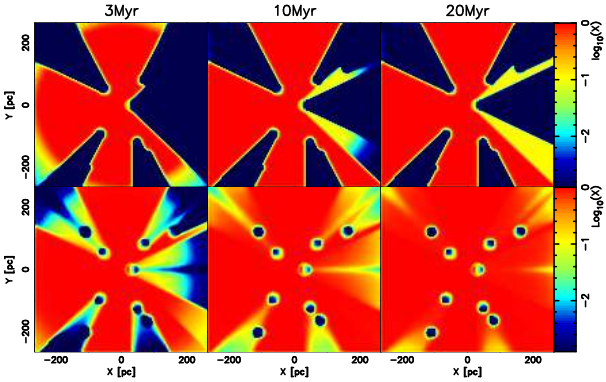}}
	\caption{Ionization of a medium containing small clumps.
The hydrogen number density and radius of each clump are
$n_{\rm c}=2.0\times 10^{-1}{\rm cm^{-3}}$ and $r_{\rm c}=10$pc, respectively.
The number density of surrounding medium is $n_{\rm H} = 10^{-2} {\rm cm^{-3}}$. 
Upper three panels are the results by the on-the-spot approximation, while
	lower three panels are the results by solving the transfer of diffuse recombination photons.
The neutral fraction in a slice through the mid-plane of the 
	computational box are shown at $t=3 {\rm Myr}$ (left panels), $t=10 {\rm Myr}$ (middle panels), 
	and $t=20 {\rm Myr}$ (right panels). }
	\label{test7Xr}
\end{figure}

We also demonstrate the effect of recombination photons 
for a medium containing multiple small clumps. 
In this simulation, nine dense clumps are set up on the mid-plane $(z=0)$. 
Each clump has the hydrogen number density of $n_{\rm c}=2.0\times 10^{-1}{\rm cm^{-3}}$
and the radius of $r_{\rm c}=10$pc.
The number density of surrounding medium is $n_{\rm H} = 10^{-2} {\rm cm^{-3}}$. 
These clumps are illuminated by a UV source, 
which is located at the box centre. 
The radius of each clump is be comparable with the mean free path of ionizing photons. 
Fig \ref{test7Xr} shows the ionization structure under the on-the-spot 
approximation (upper three panels), and that resulting from 
the transfer of recombination photons (lower three panels).  
With the on-the-spot approximation, shadows by dense clumps are crearly created. 
In this case, UV radiation never reaches shadowed regions behind the clumps. 
In contrast, if the transfer of recombination photons is solved, low density 
regions behind all clumps are highly ionized. 
Consequently, the ionization structure is dramatically changed,
compared to the on-the-spot approximation. 

\subsection{Computational time for the transfer of diffuse photons}

\begin{figure}
	\centering
	{\includegraphics[width=8cm]{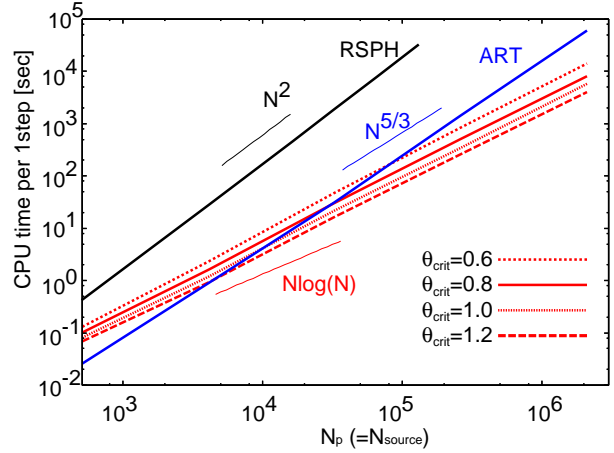}}
	\caption{Computational time for solving the transfer of diffuse recombination photons 
 	is shown as a function of 
	the particle (or grid cell) number. The computational time with START, RSPH, and 
	ART (a grid-based accelerated radiative transfer) are plotted by red, 
	black, and blue lines, respectively. As for START, the computational time with 
	$\theta_{\rm crit}=0.6$, $0.8$, $1.0$, and $1.2$ are indicated by short-dashed, 
	solid, dotted, and long-dashed lines, respectively.}
	\label{time2}
\end{figure}

In Fig.\ref{time2}, we show the average computational time per step 
for the simulations in Test 5 as a function of the number of SPH particles. 
The calculations are performed until $t=100$Myr with $\sim 10-30$ time steps. 
In this figure, RSPH, START, and ART (grid-based accelerated radiative transfer)
 are compared. 
As mentioned above, in RSPH ray-tracing method, 
the transfer of ionizing photons from each SPH particle 
to all other SPH particles are solved. 
Therefore, the computational time should follow equation (\ref{trsph}). 
On the other hand, in START scheme, distant radiation sources are regarded as one virtual bright source. 
Hence, the calculation time should follow equation (\ref{trsph2}). 
Note that the density distribution is nearly uniform at the initial phase, 
but becomes very inhomogeneous at the end of the calculation as shown in Fig. \ref{test5p}.  
Nonetheless, the measured computational time follows equation (\ref{trsph2}), as seen in Fig. \ref{time2}. 
It turns out that START is 100-1000 times faster than RSPH in the case of $N_{\rm p}\sim 10^5$. 
More dramatic speed-up is expected in simulations with larger number of particles, 
although the degree of speed-up would be different according to the distributions of particles. 
As shown in Fig.\ref{time2}, the larger $\theta_{\rm crit}$ is, the shorter the calculation time is. 
The computational time is roughly scaled to $\theta_{\rm crit}^{-2}$. 

Also, the dependence of computational time for START is better than that 
for a grid-based ray-tracing method, e.g., a long characteristics 
method or a short-characteristics method. The details are given in Appendix C.
ART is a grid-based accelerated radiative transfer scheme.
The accuracy of ART is close to a long-characteristics method, 
and the computational cost is similar to a short characteristics method \citep{Iliev06}.
The computational time with ART is basically proportional to $N_{\rm grid}^{5/3}$.
We also show measured calculation time of ART in Fig. \ref{time2}. 
Actually, the calculation time of ART is roughly proportional to ${N_{\rm grid}}^{5/3}$. 
As shown in Fig.\ref{time2}, START is faster than ART, 
if the total particle (grid) number is greater than $\sim 10^4$. 

\section{Conclusions and Discussion}

We have presented a novel accelerated radiation hydrodynamics scheme START,
which is an SPH scheme with tree-based acceleration of radiation transfer.
In this scheme, an oct-tree structure is utilized to reduce the effective number 
of radiation sources. 
The computational time by START is roughly proportional to $N_{\rm p}\log N_{\rm s}$,
where $\log N_{\rm s}$ is the radiation source number and $N_{\rm p}$ is the SPH
particle number.
We have shown the accuracy of START is almost equivalent
to that of a previous radiation SPH scheme, if we set the tolerance parameter
as $\theta_{\rm crit}\le1.0$.
The method presented in this paper provides a powerful tool to solve 
numerous radiation sources. 
Also, the new scheme allows us to solve the transfer of diffuse scattering photons.
In this case, the computational time is proportional to $N_{\rm p}\log N_{\rm p}$.  
We have simulated the expansion of an HII region around a radiation source
in an initially uniform medium. 
Comparing these results with 1D spherical symmetric RHD, we have confirmed 
that our new method correctly solve the transfer of diffuse recombination photons. 
In addition, we have explored the impacts of recombination photons in the case 
that dense clumps are irradiated by a luminous radiation source. 
As a result, we have found that the recombination photons can dramatically change 
the ionization structure especially behind clumps, if the size of the clump is 
smaller than the mean free path of the recombination photons. 

START is likely to be available for various astrophysical issues in which numerous radiation sources are required. 
One of them is the simulation of cosmic reionization. 
Many simulations have hitherto been widely performed [e.g.,  \cite{Baek09}, \cite{Iliev07}, \cite{Mellema06}, \cite{Santos08}, and \cite{Thomas09}]. 
With START, it would be possible to trace the reionization process by solving hydrodynamics consistently coupled with the transfer of UV photons from numerous sources. 
Furthermore, the diffuse radiation may play an important role when we consider the impacts of UV radiation from a massive star on neighboring medium \citep{AS07, Whalen08, HUS09, HUK09}. 
We will further challenge these issues in forth-coming papers.

\section*{Acknowledgements}
We are grateful to T. Kitayama for providing the 1D RHD code,
and H. Susa for valuable discussion. Also, we thank the anonymous
referee for useful comments.
Numerical simulations have been performed with FIRST and T2K 
at Centre for Computational Sciences in University of Tsukuba. 
This work was supported in part by the FIRST project based onGrants-in-Aid 
for Specially Promoted Research by MEXT (16002003) and Grant-in-Aid 
for Scientific Research (S) by JSPS  (20224002).

\appendix
\section{Non-equilibrium chemistry}

The non-equilibrium chemistry for $\rm e^-$, p, $\rm H$, 
$\rm H^-$, $\rm H_2$, and $\rm H^+_2$ is implicitly solved by the chemical network solver in \cite{Kitayama01}. 
Using the evaluated optical depth and omitting the suffix $i$, the photoionization rate of hydrogen for each SPH particle $i$ is
given by 
\begin{equation}
	k_{{\rm ion}} = \sum_{\alpha} k_{{\rm ion},\alpha}, 
\end{equation} 
where $k_{{\rm ion}, \alpha}$ denotes the radiative contribution from radiation source $\alpha$ on 
each particle $i$, which is represented by 
\begin{equation}
	k_{{\rm ion}, \alpha} = n_{{\rm HI}}\int^{\infty}_{\nu_{\rm L}}\int
	\frac{I_{\rm \nu, \alpha}\exp({-\tau_{\nu, \alpha}})}{h\nu}\sigma_{\nu}d\Omega d\nu, 
\end{equation}
where $n_{{\rm HI}}$, $\nu_{\rm L}$, $I_{\nu, \alpha}$, and $\sigma_{\nu}$ are 
the neutral hydrogen number density of the particle $i$, the Lyman limit frequency, 
the intrinsic specific intensity emitted from the radiation source $\alpha$, and 
the photoionization cross-section at a frequency $\nu$. 
Similarly, the photoheating rate for each SPH particle $i$ is denoted by 
\begin{equation}
	\Gamma_{{\rm ion}}= \sum_{\alpha} \Gamma_{{\rm ion},\alpha}, 
\end{equation} 
where
\begin{equation}
	\begin{array}{ccc}
	\Gamma_{\rm ion, \alpha}  = n_{\rm HI} \int^{\infty}_{\nu_{\rm L}}\int
	\frac{I_{\rm \nu, \alpha}\exp({-\tau_{\nu,\alpha}})}{h\nu} \\
	\times
	(h\nu - h\nu_{\rm L} )\sigma_{\nu}d\Omega d\nu. 
	\end{array}
\end{equation}
Here the integral with respect to solid angles is done by taking into account the effect of geometrical dilution. 
The photodissociation rate for each SPH particle $i$ is also given by 
\begin{equation}
	k_{\rm dis} = \sum_{\alpha} k_{\rm dis,\alpha},
\end{equation} 
where $k_{\rm dis,\alpha}$ is the contribution from radiation source $\alpha$, 
which is evaluated by using equation (\ref{H2ss}).

\section{H$_2$ Photo-dissociation}

$\rm H_2$ photo-dissociation is regulated by the transfer of 
Lyman-Werner (LW) band lines of $\rm H_2$ molecules. 
The transfer of LW band is explored in detail by \cite{DB96},
and the self-shielding function is provided. Here, we employ the
self-shielding function and evaluate the photo-dissociation rate as 
\begin{equation}
	k_{\rm dis} = 1.13\times 10^8 F_{\rm LW,0} f_{\rm sh}\left(\frac{N_{\rm H_2}}
	{10^{14} {\rm cm^{-2}}}\right) {\rm s^{-1}},  
	\label{H2ss}
\end{equation}
where 
\begin{equation}
	f_{\rm sh}(x) \equiv \left\{
	\begin{array}{cc}
	1,~~~~~~~~~~~~~~~~~~~~~~~~~~~~~~~~~ x \le 1 &\\
	x^{-3/4}.~~~~~~~~~~~~~~~~~~~~~~~~~~~~x > 1  &
	\end{array}
	\right. 
\end{equation}
Here $F_{\rm LW,0}$ is the LW flux in absence of the self-shielding effect 
and $N_{\rm H_2}$ is the $\rm H_2$ column density. 
The ${\rm H_2}$ column density from a radiation source to the target particle is given by 
\begin{equation}
	N_{\rm H_2, tar} = N_{\rm H_2, up} + \Delta N_{\rm H_2},
	\label{NH2}
\end{equation}
where $N_{\rm H_2,up}$ is the column density from the radiation source to the upstream particle, 
and $\Delta H_{\rm H_2}$ is given by 
\begin{equation}
	\Delta N_{\rm H_2} = \Delta l \left(\frac{n_{\rm H_2, up} + n_{\rm H_2, tar}}{2}\right). 
\end{equation}
Here $n_{\rm H_2,up}$ and $n_{\rm H_2,tar}$ are the $\rm H_2$ number densities at the points 
of the upstream particle and the target particle, respectively.

\section{Grid-based radiative transfer}

In the case where all grids emit radiation, the specific intensity of each grid cell 
is evaluated by solving equation (\ref{solution1}) along characteristics. 

In a long-characteristics method, the specific intensity at each grid point are 
evaluated for all angular directions. 
Moreover, calculations of the order of $N_{\rm grid}^{1/3}$ are needed 
to solve the equation (\ref{solution1}) along each light ray. 
Therefore, the calculation time is roughly proportional to 
$N_{\rm grid}^{4/3} \times N_{\theta} \times N_{\phi}$, where 
$N_{\phi}$, $N_{\theta}$ are the number of bins for $\phi$ direction and $\theta$ direction, respectively. 
$N_{\theta}\sim N_{\phi}$ should be of the order of $\sim N_{\rm grid}^{1/3}$, 
since all the points in the simulation box should be irradiated by all grid points. 
As a result, the computaional time is proportional to $N_{\rm grid}^2$. 

In a short characteristics method, the optical depth is calculated by interpolating 
values of segments lying along the light ray. 
When the transfer of recombination photons is involved, the total number 
of the incident rays into a computational box is 
$N_{\rm grid}^{2/3} \times N_{\theta}\times N_{\theta}\sim N_{\rm grid}^{4/3}$ \citep{Nakamoto01}. 
In addition, calculation of the order of $N_{\rm grid}^{1/3}$ are needed for each incident ray. 
Therefore, the computational time of the short characteristics method is 
proportional to $N_{\rm grid}^{5/3}$. 

Authentic Radiative Transfer (ART) scheme is based on a long-characteristics method,
but only parallel right rays are solved. As a result, the accuracy is close to 
a long-characteristics method, and the computational cost is similar to 
a short characteristics method \citep{Iliev06}.
ART scheme allows us to solve the transfer of diffuse radiation. 
Actually, the scheme is applied to evaluate the escape fraction of ionizing photons 
from a high-$z$ galaxy \citep{Yajima09}. 
In the ART scheme for a simulation of diffuse radiation, the specific intensity on 
each segment along a light ray is calculated by using equation (\ref{solution1}). 
The intesity at a grid point is obtained by the interpolation from the neighbouring
light rays. The total number of the light rays  is 
$6 \times N_{\rm grid}^{2/3}\times N_{\theta}\times N_{\theta}\sim N_{\rm grid}^{4/3}$. 
Thus, the calculation time for this method is proportional to $N_{\rm grid}^{5/3}$, 
as similar to that for a short characteristics method.

\label{lastpage}

\end{document}